\documentclass{aa}
\usepackage{graphicx}
\begin{document}
\title{Observation of coronal loop torsional oscillation}

\author{T.V. Zaqarashvili}
\offprints{T. Zaqarashvili}

\institute{ Abastumani Astrophysical Observatory, Al. Kazbegi ave.
2a, 380060 Tbilisi, Georgia \\ \email{temury@mcs.st-and.ac.uk}}

\date{Received November 22, 2002/ Accepted January 14, 2003}

   \abstract{
We suggest that the global torsional oscillation of solar coronal
loop may be observed by the periodical variation of a spectral
line width. The amplitude of the variation must be maximal at the
velocity antinodes and minimal at the nodes of the torsional
oscillation. Then the spectroscopic observation as a time series
at different heights above the active region at the solar limb may
allow to determine the period and wavelength of global torsional
oscillation and consequently the Alfv{\'e}n speed in corona. From
the analysis of early observation (Egan \& Schneeberger
\cite{egan}) we suggest the value of coronal Alfv{\'e}n speed as
$\sim 500 \,{\rm km}{\cdot}{\rm s}^{-1}$.
   }

\maketitle

\section{Introduction}

Magnetohydrodynamic (MHD) waves and oscillations play an important
role in the dynamics of the solar corona (Roberts
\cite{roberts2}). They can heat and accelerate the coronal plasma
(Goossens et al. \cite{goossens}) and also offer an unique tool
for developing a coronal seismology (Edwin \& Roberts
\cite{edwin}, Nakariakov \& Ofman \cite{nakariakov2}). The solar
corona is highly structured into narrow coronal magnetic loops
which are anchored into the dense photosphere. The loops permit
the propagation of three kinds of MHD waves: sausage, kink and
torsional Alfv{\'e}n waves. The closed boundary conditions at the
loop footpoints lead to the discrete spectrum of harmonics and
consequently to the formation of global oscillations.

The global MHD oscillation in the sausage mode (Roberts et al.
\cite{roberts1}) modulate loop cross section and causes the
variation of the density and emission measure of radiating plasma.
Using coronal loop images obtained with the Yohkoh soft X-ray
telescope (SXT), McKenzie \& Mullan (\cite{mckenzie}) found the
periodical modulation of the X-ray brightness, which was
interpreted as the signature of global mode oscillation. The
global kink oscillations can be detected as transverse
displacement of spatial loop positions between fixed nodes. The
typical property of recently observed kink oscillations by TRACE
(Aschwanden et al. \cite{aschwanden}, Nakariakov et al.
\cite{nakariakov1}) is very short damping time. It was suggested
that the oscillations may be damped either due to the phase mixing
with anomalously high viscosity (Ofman \& Aschwanden
\cite{ofman1}) or due to the resonant absorption (Ruderman \&
Roberts \cite{ruderman}) acting in the inhomogeneous regions of
the tube, which leads to a transfer of energy from the kink mode
to Alfv{\'e}n (azimuthal) oscillations within the inhomogeneous
layer. Consequently the resonant absorption of global kink modes
may lead to the global torsional oscillation of coronal loops.

Unfortunately, contrary to the sausage and kink modes, the global
torsional oscillations do not cause neither intensity variation
nor the spatial displacement of the loop position. Therefore they
hardly subject to the observation in usual coronal spectral lines.

The influence of torsional waves propagating along a thin,
vertical, photospheric flux tube on Zeeman-split polarized line
profiles (Stokes profiles) was studied recently by Ploner \&
Solanki (\cite{ploner}). In the presence of such a wave spatially
resolved Stokes profiles are found to oscillate strongly in
wavelength, amplitude and blue-red asymmetry. Also observation on
the Solar and Heliospheric Observatory (SOHO) showed an increase
of spectral line width with height (Doyle et al. \cite{doyle},
Banerjee et al. \cite{banerjee}) which can be interpreted as
vertically propagating undamped Alfv{\'e}n waves (Moran
\cite{moran}). However no wave parameters (wavelength, frequency)
were identified, because of the propagating pattern of waves. On
the other hand, the stationary character of standing Alfv{\'e}n
waves in magnetic tubes (or torsional oscillations) allows to
identify the velocity nodes and antinodes and thus may be used for
determining the wave parameters.

We suggest that the global torsional oscillation of coronal loop
can be detected by periodical broadening of spectral lines due to
the periodic azimuthal velocity. The amplitude of the azimuthal
velocity in the torsional oscillation is maximal at the antinodes
and tends to zero at the nodes. Therefore the torsional
oscillation causes the periodical variation of spectral line
broadening with different amplitudes at different heights from the
solar surface: the place of stronger variation corresponds to the
antinode and the place of constant broadening corresponds to the
node. Then the time series of spectroscopic observations at
different heights above the active region at the solar limb may
allow to determine the period and wavelength of global torsional
oscillation and consequently the Alfv{\'e}n speed in the coronal
loops.

Recently, strongly damped Doppler shift oscillations of hot (T$>$6
MK) coronal loops were observed with SUMER on SOHO (Kliem et al.
\cite{kliem}, Wang et al. \cite{wang}). These oscillations were
interpreted as signatures of slow-mode magnetosonic waves excited
impulsively in the loops (Ofman \& Wang \cite{ofman2}). It is
possible that the slow magnetosonic waves (as well as the kink
waves) may also cause the oscillation of Doppler broadening in
given direction of propagation. However the velocity field of
waves is inhomogeneous (which is necessary for the line
broadening) only at wave nodes and thus may lead only to
negligible effect. On the other hand, the velocity field at wave
antinodes will cause the Doppler shift oscillation as was
interpreted by Ofman \& Wang (\cite{ofman2}). However it is also
possible that the Doppler broadening of the line is due to
superimpose of several loops in the line of sight containing slow
waves at different phases. But in this case the broadening must be
nearly constant or it will show rather chaotic behaviour in time
than the periodical one. Recently Sakurai et al. (\cite{sakurai})
presented a time sequence over 80 min of coronal green-line
spectra obtained with a ground-based coronagraph at the Norikura
Solar Observatory. They found the oscillations of the spectral
line Doppler shifts and interpreted them as propagating slow MHD
waves. No clear evidence of the Doppler width oscillations was
found. It may indicate that either they could not catch the
velocity antinodes of the global torsional oscillation or the
oscillations are not common in corona and they are related to the
flare or mass ejection. Future space and ground-based
spectroscopic observations are necessary to make the conclusion.

In the next section we briefly describe the method of observation
and interpret the early observation of Eagen \& Schneeberger
(\cite{egan}) as the global torsional oscillation of coronal loop.

\section{Observation of the torsional oscillation}

Consider a coronal loop as straight magnetic tube, the axis of
which is perpendicular to the line of sight. If the tube rotates
about its axis with an angular velocity $\Omega$, then the Doppler
shift of the observed spectral line with wavelength $\lambda$ is
$\Delta{\lambda}=({{\lambda\Omega}/c})x$, where $x$ is the
distance from the tube axis in the plane perpendicular to the line
of sight and $c$ is the speed of light (the expression must be
multiplied by $\sin{i}$ if the tube is inclined with the angle
$i$). The largest shift occurs at the edges of the tube and has a
value
\begin{equation}
\Delta{\lambda}={{\lambda}\over c}v,
\end{equation}
where $v$ is the rotation velocity at $x=R$ ($R$ is the radius of
the tube). So the Doppler shifts at different $x$ from the tube
axis cause the spectral line broadening by $2\left
|\Delta\lambda\right |$. It must be mentioned that the broadening
of spectral lines can be caused by thermal motion of ions and
turbulent motion of gas, however here we consider only the
nonthermal broadening of spectral lines caused by the azimuthal
velocity of the tube.

Now suppose that the tube undergoes the torsional oscillation due
to the closed boundary conditions imposed at the tube ends. For
simplicity the density $\rho$ and the magnetic field ${\bf
B}=(0,0,B_0)$, directed along the axis $z$ of a cylindrical
coordinate system $r,\phi,z$, are considered to be homogeneous
throughout the tube. Then the velocity and the magnetic field
components of the torsional oscillation can be expressed as
\begin{equation}
B_{\phi}=-B_0{\alpha}{\cos}(\omega_nt){\cos}(k_nz),
\end{equation}
\begin{equation}
V_{\phi}=V_A{\alpha}{\sin}(\omega_nt){\sin}(k_nz),
\end{equation}
where $V_A={{B_0}/{\sqrt {4\pi{\rho}}}}$ is the Alfv{\'e}n speed,
$k_n={{n\pi}/{L}}$ ($L$ is the length of the tube and n=1,2,3...)
is eigenvalue of wave number, $\omega_n$ is the corresponding
eigenfrequency and $\alpha$ is the relative amplitude of
oscillations. Eigenvalues and eigenfrequencies satisfy the
dispersion relation of Alfv{\'e}n waves
\begin{equation}
{{\omega_n}\over {k_n}}=V_A.
\end{equation}
The velocity field of each $n$th mode has different amplitude of
oscillation, but the same frequency $\omega_n$ along the tube: the
amplitude is maximal at the antinodes $(\sin(k_nz)={\pm}1)$ and
tends to zero at the nodes $(\sin(k_nz)=0)$.

Then the velocity field (3) leads to the periodical variation of
Doppler width at each point $z$ along the tube axis. It means that
the torsional oscillation causes the periodical variation of
spectral line broadening (expressed by a half of line width,
$\Delta\lambda_B$, hereafter HW) with the same frequency, but with
different amplitudes along the tube. The amplitude of HW variation
will be maximal at the velocity antinodes and minimal at the
nodes. According to the equations (1) and (3) HW can be expressed
by
\begin{equation}
\Delta{\lambda}_B= {{{\alpha}V_A{\lambda}}\over {c}}{\left
|{\sin}(\omega_nt){\sin}(k_nz)\right |}.
\end{equation}

The distance between the points of maximal variation i.e. between
the velocity antinodes will be the half wavelength of torsional
oscillation. Then the observed period of HW variation allows to
determine the Alfv{\'e}n speed from the dispersion relation (4).
Also the amplitude of HW variation at the velocity antinodes
(maximal value of $\Delta\lambda_B$ along the tube axis) gives the
value of $\alpha$ i.e. the amplitude of torsional oscillations.

However we have to be sure that the periodical variation of
spectral line broadening can be caused only by the torsional
oscillations. The velocity field of other wave motions may cause
the Doppler shift variations. For example, the slow magnetosonic
waves propagating along an applied magnetic field may lead to this
effect recently observed by SUMER in given arbitrary loop
orientation (Kliem et al. \cite{kliem}, Wang et al. \cite{wang},
Ofman \& Wang \cite{ofman2}). However neither slow magnetosonic
waves, nor kink waves can cause the significant periodical
variation of Doppler broadening at given height from the solar
surface. The velocity field of waves must be highly inhomogeneous
to produce the line broadening. The velocity field of slow
magnetosonic waves (as well as the kink waves) is inhomogeneous at
velocity nodes, where they can produce only negligible Doppler
broadening. On the other hand, they can produce the Doppler shift
oscillations at antinodes where the velocity is maximal and almost
homogeneous. However it is also possible that the Doppler
broadening of the line is due to superimpose of several loops in
the line of sight containing slow waves at different phases. But
in this case the broadening must be nearly constant or it will
show rather chaotic behaviour in time than the periodical one.
Therefore from our point of view the torsional waves are best
candidates for producing the periodical variation of spectral line
broadening.

Thus the observation as a time series at different heights above
the active region at the solar limb may allow to determine the
wavelength and the period of torsional oscillation. As an example
of proposed method let analyse the observation made almost two
decades ago by Egan \& Schneeberger (\cite{egan}). They presented
the time series of Fe XIV coronal emission line spectra above both
active and quiet regions at the solar limb. They found the Doppler
width temporal variations with the period of $6.1\pm0.6$ min above
the active region. The width fluctuations were amplified in two
5500 km long segments at heights of 66 000 km and 88 000 km above
the limb. The amplitude of variation was 0.125 {\AA} at 88 000 km,
while at lowest heights it reduces to 0.04 {\AA}. They also found
a marginal evidence for 6 min intensity oscillations at these two
positions. For quite region they found neither Doppler width nor
intensity variations.

We suggest that this observation can be interpreted as the global
torsional oscillation of coronal loop. The two regions of
amplified Doppler width variation may correspond to the velocity
antinodes of torsional oscillation. They likely belong to the
second mode of oscillation, because it has two velocity antinodes
at the middle of both footpoints, while first mode has only one at
the top (the velocity antinode at the loop top hardly undergo to
the observation in this spectral line because of rapidly
decreasing intensity in the corona).
   \begin{figure}
   \centering
   \includegraphics[width=6cm]{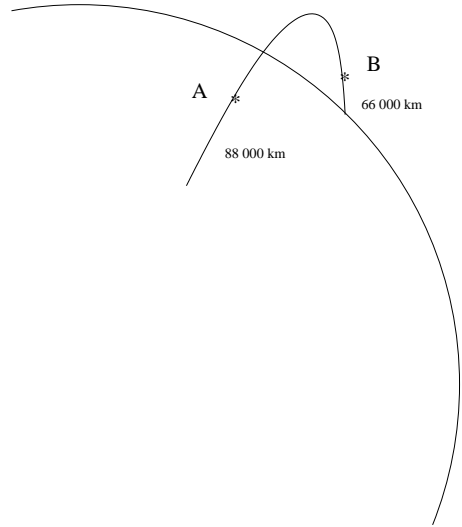}
      \caption{Schematic picture of a coronal loop at the solar limb is presented.
      The points A and B correspond to the velocity antinodes of second mode of global torsional oscillation.
      Both points are located at $\sim$ 88 000 km from the surface. However only
       66 000 km part is visible in the case of point B, because the footpoint
       is anchored into the reverse side of the solar surface.
              }
         \label{FigVibStab}
   \end{figure}

As the observations were performed off the solar limb it is
reasonable to suppose that the antinodes were located at $\sim$ 88
000 km height from the photosphere, but the second footpoint was
anchored at the reverse invisible side of the Sun, so that only 66
000 km part of footpoint was visible (Fig.1). Then the wavelength
of global torsional oscillation can be
\begin{equation}
\lambda=4{\cdot}\,88\,000 \,{\rm km}\,{\approx}\, 352\, 000\, {\rm
km}.
\end{equation}
The period of torsional oscillations will be $\sim$12.2 min i.e.
double of observed period. Then we easily calculate the Alfv{\'e}n
speed in corona (here the constant Alfv{\'e}n speed is assumed
along the loop, however in real conditions it may be inhomogeneous
which complicates the problem)
\begin{equation}
V_A \sim 500 \,{\rm km}{\cdot}{\rm s}^{-1}.
\end{equation}

If the regions of amplified Doppler width oscillations belong to
two different loops, then the similar calculation for the loop
with velocity antinodes at 66 000 km gives
\begin{equation}
V_A \sim 360 \,{\rm km}{\cdot}{\rm s}^{-1},
\end{equation}
which is below the expected value of the Alfv{\'e}n speed in
corona. Also it seems unreal that the two different loops have the
same periods of global mode. On the other hand, the global
oscillation of coronal loop in other wave mode may not cause the
periodic oscillations of Doppler width. Only the temperature
changing due to the nonadiabatic slow MHD waves may lead to this
variation. But it seems again unreal that the nonadiabaticity
leads to the variation of line broadening with $\sim25{\%}$.
Therefore we suggest that the torsional oscillation of one loop
can be the reason for the observed phenomenon.

We also easily calculate the amplitude of torsional oscillation
for Fe XIV ($\lambda5303$) spectral line from the equation (5)
\begin{equation}
\sim7\, {\rm km}{\cdot}{\rm s}^{-1}.
\end{equation}
This result is calculated when the loop axis is perpendicular to
the line of sight. The amplitude of oscillation will be even
greater in the case of loop inclination.

Evaluated Alfv{\'e}n speed, as well as the amplitude of torsional
oscillations, have the suggestive values for the low corona. This
gives the idea that the observation of Egan \& Schneeberger
(\cite{egan}) may be interpreted as the global torsional
oscillation of coronal loop. Observed intensity oscillations can
be interpreted as the density perturbations caused by the
nonlinear magnetic pressure acting at velocity antinodes of
torsional oscillations.

Another problem is that the amplitude of the Doppler width
oscillations in the paper of Egan \& Schneeberger (\cite{egan}) is
comparable or below the instrumental spectral resolution.
Therefore the validity of the observation is unclear. On the other
hand, Sakurai et al. (\cite{sakurai}) could not find clear
periodical variation of the line width. It may indicate that
either they could not catch the velocity antinodes of the global
torsional oscillation or the oscillations are not common in corona
and they are related to the flare or mass ejection like the
transverse loop oscillations (Schrijver et al. \cite{schrijver}).
Therefore the future observations are necessary for making the
final conclusion. Also analysis of recent space based
spectroscopic observations on SOHO and TRACE may reveal more
precise data which can test the validity of this method.

\section{Conclusion}

We suggest that the global torsional oscillation of the coronal
loop, axis of which is not parallel to the line of sight, can be
observed by the periodical variation of spectral line width. The
amplitude of the variation must be maximal at certain regions
which correspond to the velocity antinodes of torsional
oscillations. Then the period of variation will be the half of
oscillation period and the distance between the regions with
maximal variations will be the half of wavelength. The
spectroscopic observation as a time series at different height
above the active region at the solar limb may allow to determine
the wavelength and the period of the torsional oscillation. Then
the Alfv{\'e}n speed can be calculated which is very important for
determination of magnetic field strength in corona. The
observation of the spectral lines with larger wavelengths will be
more successful, because of their sensitivity to the Doppler
shift. Both, ground based coronagraphs and space based instruments
can be used. The analysis of early observation (Egan \&
Schneeberger \cite{egan}) give the values of Alfv{\'e}n speed and
the amplitude of torsional oscillations as $\sim 500 \,{\rm
km}{\cdot}{\rm s}^{-1}$ and $>7\, {\rm km}{\cdot}{\rm s}^{-1}$
respectively.

We hope that the observation of torsional oscillations after the
damping of coronal loop kink oscillations will provide the clue
for the damping mechanism and consequently for the coronal heating
(Ruderman \& Roberts \cite{ruderman}).

\begin{acknowledgements}
The author thanks the anonymous referee for helpful suggestions.
\end{acknowledgements}

\end{document}